\documentclass[preprint,aps,showpacs]{revtex4}
\usepackage[dvips]{graphicx}
\usepackage{amsmath,amssymb}

\newcommand{\bnu}{\mbox{\boldmath$\nu$}}
\newcommand{\be}{\mbox{\boldmath$e$}}
\newcommand{\bzero}{\mbox{\boldmath$0$}}

\begin{document}

\title{
Systematic generation of entanglement measures 
for pure states
}
\author{Ayumu Sugita}
\email{sugita@a-phys.eng.osaka-cu.ac.jp}
\affiliation{
Department of Applied Physics, Osaka City University
3-3-138 Sugimoto, Sumiyoshi-ku, Osaka, 558-8585, Japan}

\date{\today}
            
\begin{abstract}
We propose a method to generate entanglement measures systematically
by using the irreducible decomposition of
some copies of a state under the local unitary 
(LU) transformations. It is applicable to general multipartite systems.
We show that there are entanglement monotones corresponding
to singlet representations of the LU group. They can be evaluated
efficiently in an algebraic way,
and experimentally measurable by local projective
measurements of the copies of the state.
Non-singlet representations
are also shown to be useful to classify entanglement.
Our method reproduces many well-known measures
in a unified way, and produces also a lot of new ones.

\end{abstract}
            
\maketitle  
 
\section{Introduction}
Entanglement is one of the most striking features of quantum
mechanics, and considered to be the key resource for
quantum-information processing.  In spite of the intensive
study in the last decade, description and quantification of 
multipartite entanglement is still a challenging problem.
A single quantity is not enough to characterize entanglement
of multipartite states, since there can be qualitatively different
quantum correlations.
(A famous example is the Greenberger-Horne-Zeilinger (GHZ) state and the
W state in 3-qubit systems \cite{3qubit}.)  
Therefore we need more than one 
measure to classify multipartite entanglement, and the
number of the necessary measures grows rapidly
as the number of parties increases.

An important requirement for entanglement measure is
to be entanglement monotone, i.e., nonincreasing
under stochastic local operations and classical
communication (SLOCC). So far, constructions of entanglement monotones 
have been made mostly in heuristic ways using specific
features of the system under consideration.
To the best of the author's knowledge, the only method
of construction of entanglement monotones which is applicable
to general multipartite system is the
hyperdeterminant \cite{miyake}.
However, we can make only one entanglement monotone for a 
system from the hyperdeterminant. Furthermore,
the polynomial degree of the hyperdeterminant grows very fast 
as the number of the parties increases,
which makes it difficult to write down its explicit form.
Therefore we need a method to generate many entanglement monotones
for an arbitrary multipartite system, hopefully in a systematic and efficient
way.

In this paper, we propose a method which satisfies all these
requirements.
We consider $q$ copies of a state $|\psi^{\otimes q}\rangle$,
where $q$ is an arbitrary positive integer, and decompose it
into irreducible components under the LU group. 
Then the norm of an irreducible
component can be regarded as a measure of entanglement.
Its explicit form is given by the Clebsch-Gordan coefficients
of the LU group and the expansion coefficients of the given state.
Our method enables us to make a list of all algebraic invariants of the LU 
group of a given order.

Irreducible decomposition of $|\psi^{\otimes q}\rangle$ 
was first introduced in \cite{sugita1} in the context of 
algebraic evaluation of the moments of the generalized
Husimi distribution. 
Our method can be regarded as
a generalization of the definition of the 
concurrence for multipartite systems
in \cite{concurrence2}, in which irreducible decomposition
of $|\psi^{\otimes 2}\rangle$ was considered. Our method
also includes the construction of entanglement
monotones for multiqubit states in \cite{osterloh}
with use of expectation values of antilinear operators.
The expectation value of an antilinear operator $A$ is
written as $\langle A\rangle =\langle \psi|LC|\psi\rangle$, 
where $L$ is a linear operator and $C$ is the complex conjugation.
If we expand the state as $|\psi\rangle = \sum_j c_j |j\rangle$,
the expectation value can be written explicitly as $\sum_{i,j}
= L_{i,j} c_i^* c_j^*$, where $L_{i,j} = \langle i|L|j\rangle$.
Then $\langle A\rangle^* = \sum_{i,j} L_{i,j}c_i c_j$ is considered as a linear map from 
$|\psi^{\otimes 2} \rangle$ to a complex number. Therefore
if $| \langle A\rangle |$ is invariant under the LU group, it must be
obtained from a singlet component of $|\psi^{\otimes 2}\rangle$.
\footnote{The first and the second order "combs" in 
\cite{osterloh} correspond to $|s_2\rangle$ and $|s_{4,a}\rangle$, respectively, in this paper.}

The purpose of this paper is to show that our group theoretical method 
gives a quite general and unified point of view for 
multipartite entanglement measures. Therefore, in the following,
we concentrate mainly on describing the general theory and deriving 
well-known existing measures from our method.
Analysis of complicated entanglement in specific systems
will be reported elsewhere.

\section{Description of the general method}
First we describe our method as generally as possible. 
Let us consider an $m$-partite system
with $N_i$ states for $i$th party.
The LU group of this system is ${\rm SU}(N_1)\times {\rm SU}(N_2) 
\times \dots \times {\rm SU}(N_m)$.
Since an irreducible representation (irrep) of a special unitary group
is specified by its highest weight \cite{georgi}, we denote an irrep with the 
highest weight $\nu$ by $R_\nu$. 
Then an irreducible representation of the LU group can be written as 
$R_{\nu_1}\otimes R_{\nu_2}\otimes \dots \otimes R_{\nu_m}$, where
$R_{\nu_i}$ is the irreducible representation of ${\rm SU}(N_i)$
with the highest weight $\nu_i$. We denote this representation
by $R_{\bnu}$, where the bold symbol $\bnu \equiv (\nu_1, \nu_2,\dots,\nu_m)$
represents the set of the highest weights.  
A pure state $|\psi\rangle$ in this system is in the defining representation
of the LU group, which is denoted by $R_{\be}$. Here, $\be \equiv (e_1,e_2,\dots,e_m)$,
and $e_i$ is the highest weight of the defining 
representation of ${\rm SU}(N_i)$. 
Explicit form of the highest weight depends on the choice of the basis of the Lie
algebra. For example, in the convention of Ref. \cite{georgi}, 
$e_i = \left(\frac{1}{2}, \frac{1}{2\sqrt{3}},\dots, 
\frac{1}{\sqrt{2m(m+1)}},\dots, \frac{1}{\sqrt{2(N_i-1)N_i}}\right)$.

Then we consider $q$ copies of a state 
$|\psi^{\otimes q}\rangle \equiv
|\psi\rangle^{\otimes q}$. Since $R_{\be}^{\otimes q}$ is reducible
for $q\ge 2$, we can decompose it into 
irreducible components as
\begin{eqnarray}
R_{\be}^{\otimes q} = \oplus_\alpha R_{\bnu_\alpha}.
\end{eqnarray}
$|\psi^{\otimes q}\rangle$ is decomposed correspondingly as
\begin{eqnarray}
|\phi^{\otimes q}\rangle = \sum_\alpha P_{\bnu_\alpha} |\phi^{\otimes q}\rangle,
\end{eqnarray}
where $P_{\bnu_\alpha}\equiv P_{\nu_{\alpha,1}}\otimes \dots \otimes P_{\nu_{\alpha,m}}$ 
is the projection operator to the representation
space of $R_{\bnu_\alpha}$. 
Among the irreducible components, there is
always a unique component with the "maximum" highest weight $q\be = (qe_1,\dots,qe_m)$.
It has been shown in \cite{sugita2} that $|\psi^{\otimes q}\rangle \in
R_{q\be}$ iff $|\psi\rangle$ is not entangled.
Therefore, for $q\ge 2$, we can conclude that
\begin{eqnarray*}
\mbox{$|\psi\rangle$ is unentangled}\;\;\;
\Longleftrightarrow \;\;\;
\left| P_{q\be} |\psi^{\otimes q}\rangle \right| = 1
\end{eqnarray*}
assuming $|\psi\rangle$ is normalized, and 
\begin{eqnarray*}
\mbox{$|\psi\rangle$ is unentangled}
\;\;\;
\Longrightarrow \;\;\;
P_{\bnu_\alpha} |\psi^{\otimes q}\rangle = 0
\end{eqnarray*}
for $\bnu_\alpha \ne q\be$.

It has been shown that 
$\left| P_{q\be} |\psi^{\otimes q}\rangle \right|^2$ is the $q$-th
moment of a generalized Husimi distribution up to a constant multiplier \cite{sugita1, sugita2},
and the R\'enyi subentropy defined from the moment is 
an entanglement monotone for $N\times N$ bipartite systems
\cite{mintert_zyczkowski}.

$\left| P_{\bnu_\alpha} |\psi^{\otimes q}\rangle \right|$ with $\bnu_\alpha \ne q\be$
is also a good candidate measure of entanglement since it is invariant 
under LU transformations and vanishes for unentangled states. 
Singlet representations, which have the "minimum" highest weight
$\bnu_\alpha = \bzero \equiv (0,\dots 0)$, are of particular interest. 
(Note that $0$ within the bracket is not a number, but the zero weight vector.)
Actually we can show
that $\left|P_{\bnu_\alpha = \bzero} |\psi^{\otimes q}\rangle \right|^{1/q}$
is an entanglement monotone.

{\em Proof of monotonicity:}
We use a theorem proved in \cite{verstraete}
which claims that {\it a linearly homogeneous positive function of a pure
state that remains invariant under determinant 1 SLOCC operations
is an entanglement monotone.} Since 
it is obvious that 
$\left|P_{\bnu_\alpha = \bzero} |\psi^{\otimes q}\rangle \right|^{1/q}$
is linearly homogeneous and positive, what we have to show is
its 
invariance under 
the SLOCC group $SL(N_1,\mathbb{C})\times \dots \times SL(N_m,\mathbb{C})$.

If we consider a singlet representation $R_{\nu=0}$
of ${\rm SU}(N)$, it is obvious that $T|\phi\rangle = 0$ for
any $T \in {\mathfrak su}(N)$ and
$|\phi\rangle \in R_{\nu=0}$,
where the fracture letters denote the Lie algebra
of the corresponding Lie group.
Since ${\mathfrak sl}(N,\mathbb{C})$ is obtained as the complexification
of ${\mathfrak su}(N)$, i.e., the set of linear combinations
of its elements with complex coefficients, $T|\phi\rangle = 0$
holds also for $\forall\, T\in {\mathfrak sl}(N,\mathbb{C})$. 
Therefore an element of the singlet representation of
${\rm SU}(N)$ is also invariant under $SL(N,\mathbb{C})$.
In the same way, $P_{\bnu_\alpha=\bzero} |\psi^{\otimes q}\rangle$
is shown to be invariant under $SL(N_1, \mathbb{C})\times \dots \times SL(N_m, \mathbb{C})$. 

\section{ 2-qubit case}
Let us consider a 2-qubit system to see how our general theory works. 
A qubit is in the defining (spin 1/2) representation of ${\rm SU}(2)$, which is $R_{1/2}$ 
in our notation. Note that the highest weight is
the total spin quantum number in this case.
In general, the tensor product of two irreducible representations
of ${\rm SU}(2)$ is decomposed as $R_{m}\otimes R_{n}
= R_{m+n}\oplus R_{m+n-1}\oplus \dots \oplus R_{|m-n|}$.

In the simplest case $q=2$, two copies of a qubit
is decomposed into a triplet $R_{1}$ and a singlet
$R_{0}$. Hereafter we arrange the tensor product
to represent copies of a quantum state vertically
in order to distinguish it from the tensor product to represent
multipartite states. Then the irreducible decomposition
of the two copies of a qubit can be written as
\begin{eqnarray}
R_{1/2}^{\otimes 2}
&=& 
\begin{array}{c}
R_{1/2}\\
\otimes \\
R_{1/2} 
\end{array}
=
R_{1} \oplus R_{0}.
\end{eqnarray} 
Since a 2-qubit state is in
the representation $R_{1/2}\otimes R_{1/2}$,
two copies of a 2-qubit state is decomposed as
\begin{eqnarray}
\begin{array}{c}
(R_{1/2}\otimes R_{1/2}) \\
\otimes \\
(R_{1/2}\otimes R_{1/2})
\end{array}
&=&
\left(
\begin{array}{c}
R_{1/2} \\ \otimes \\ R_{1/2}
\end{array}
\right)
\otimes
\left(
\begin{array}{c}
R_{1/2} \\ \otimes \\ R_{1/2}
\end{array}
\right)
\\
&=& 
(R_{1} \oplus R_{0})\otimes
(R_{1} \oplus R_{0})
\\
&=&
R_{1}\otimes R_{1} \;\oplus\; R_{1}\otimes R_{0}
\;\oplus\;
R_{0}\otimes R_{1} \;\oplus\; R_{0}\otimes R_{0}.
\end{eqnarray}
Here, $R_{1}\otimes R_{0}$ and $R_{0}\otimes R_{1}$ are antisymmetric
with respect to the exchange of the
two copies, because $R_{1}$ is symmetric and $R_{0}$ is antisymmetric.
Since $|\psi^{\otimes 2}\rangle$ is symmetric, $P_{1}\otimes P_{0}|\psi^{\otimes 2}\rangle$
and $P_{0}\otimes P_{1}|\psi^{\otimes 2}\rangle$ vanish identically.
Hence the irreducible decomposition of the two copies of a 2-qubit state is
\begin{eqnarray}
|\psi^{\otimes 2}\rangle &=&
P_{1}\otimes P_{1}|\psi^{\otimes 2}\rangle
+
P_{0}\otimes P_{0}|\psi^{\otimes 2}\rangle.
\end{eqnarray}
The squared norm of the first term is the second moment of the 
generalized Husimi distribution \cite{sugita1, gert}
up to a constant multiplier.
The second term is the projection to the singlet representation $R_{0}\otimes R_{0}$.
Hence the square root of its norm is an entanglement monotone.

Let us derive the explicit form of $P_{0}\otimes P_{0}|\psi^{\otimes 2}\rangle$.
The basis vector of the second order singlet $R_{0}$ is
\begin{eqnarray}
|s_2\rangle 
&\equiv&
\frac{1}{\sqrt{2}}
\left(
\left|
\begin{array}{c}
0 \\ 1
\end{array}
\right\rangle
-
\left|
\begin{array}{c}
1 \\ 0
\end{array}
\right\rangle
\right)\\
&=&
\frac{\epsilon_{ij}}{\sqrt{2}}
\left|
\begin{array}{c}
i \\ j
\end{array}
\right\rangle,
\end{eqnarray}
where $\epsilon_{ij}$ is the completely antisymmetric tensor. 
Note that the Einstein summation convention is used and
the indices take on the values $0$ or $1$.
Then the basis vector of $R_{0}\otimes R_{0}$ is
\begin{eqnarray}
&&
|s_2\rangle \otimes |s_2\rangle \\
&=&
\frac{1}{2}
\epsilon_{i_{11}i_{21}} \epsilon_{i_{12}i_{22}}
\left|
\begin{array}{cc}
i_{11} & i_{12} \\
i_{21} & i_{22}
\end{array}
\right\rangle
\\
&=& 
\frac{1}{2}
\left(
\left|
\begin{array}{cc}
0 & 0 \\
1 & 1
\end{array}
\right\rangle
+
\left|
\begin{array}{cc}
1 & 1 \\
0 & 0
\end{array}
\right\rangle
-
\left|
\begin{array}{cc}
0 & 1 \\
1 & 0
\end{array}
\right\rangle
-
\left|
\begin{array}{cc}
1 & 0 \\
0 & 1
\end{array}
\right\rangle
\right).
\end{eqnarray}
Note that the set of indices $\{i_{jk}\}$ forms a $q\times m$ matrix in general. 
We expand the state $|\psi\rangle$ in the standard basis as
$
|\psi\rangle = c_{ij}|ij\rangle
$.
Then
\begin{eqnarray}
|\psi^{\otimes 2}\rangle 
&=& 
\begin{array}{c}
c_{ij}|ij\rangle \\
\otimes \\
c_{kl}|kl\rangle
\end{array}
\\
&=&
c_{i_{11}i_{12}}c_{i_{21}i_{22}}
\left|
\begin{array}{cc}
i_{11} & i_{12} \\
i_{21} & i_{22}
\end{array}
\right\rangle .
\end{eqnarray}
Therefore
\begin{eqnarray}
\left|P_{0}\otimes P_{0}|\psi^{\otimes 2}\rangle\right|
&=&
\left|\bigl(\langle s_2|\otimes \langle s_2|\bigr) |\psi^{\otimes 2}\rangle\right|\\
&=&
\frac{1}{2}
\left|
\epsilon_{i_{11}i_{21}} \epsilon_{i_{12}i_{22}}
c_{i_{11}i_{12}}c_{i_{21}i_{22}}
\right|
\\
&=&
\left|c_{11}c_{00} - c_{10}c_{01}\right|.
\end{eqnarray}
This is the concurrence for pure states except for a factor of two.

\section{3-qubit case}
Next we consider a 3-qubit system. 
For $q=2$, we have a singlet $|s_2\rangle\otimes |s_2\rangle \otimes |s_2\rangle$
but this is antisymmetric with respect to the exchange of the two copies.
Therefore the projection of $|\psi^{\otimes 2}\rangle$, where $|\psi\rangle$ is a 3-qubit state, 
to this component vanishes identically. 

Then we have to consider larger $q$ to find 
nontrivial entanglement monotones. Since it is impossible to make a singlet from 
three copies a qubit, the next candidate is $q=4$. In this case, the irreducible decomposition is
\begin{eqnarray}
R_{1/2}^{\otimes 4} &=& \left(R_{1}\oplus R_{0}\right)^{\otimes 2} =
R_{2} \oplus 3 R_{1} \oplus 2 R_{0}.
\end{eqnarray}
Therefore 
we can make two different singlets.
One is included in the combination of the two triplets $R_{1}^{\otimes 2}
= R_{2}\oplus R_{1} \oplus R_{0}$.
Its explicit form is 
\begin{eqnarray}
|s_{4,a}\rangle &\equiv & 
\frac{1}{2\sqrt{3}}\left(
2
\left|
\begin{array}{c}
0 \\ 0 \\ 1 \\ 1
\end{array}
\right\rangle
+
2 
\left|
\begin{array}{c}
1 \\ 1 \\ 0\\ 0
\end{array}
\right\rangle
-
\left|
\begin{array}{c}
0 \\ 1 \\ 0\\ 1
\end{array}
\right\rangle
-
\left|
\begin{array}{c}
1 \\ 0 \\ 1\\ 0
\end{array}
\right\rangle
-
\left|
\begin{array}{c}
0 \\ 1 \\ 1\\ 0
\end{array}
\right\rangle
-
\left|
\begin{array}{c}
1 \\ 0 \\ 0\\ 1
\end{array}
\right\rangle
\right) \\
&=&
d_{ijkl}
\left|
\begin{array}{c}
i \\ j \\ k\\ l
\end{array}
\right\rangle,
\end{eqnarray}
where
\begin{equation}
d_{ijkl} \equiv 
\left\{
\begin{array}{ll}
 \frac{1}{\sqrt{3}} & \left[ (ijkl) = (1100), (0011)\right]\\
-\frac{1}{2\sqrt{3}}  & \left[(ijkl) = (1010), (0101), (1001),(0110) \right]\\
 0 & \mbox{else}
\end{array}
\right. .
\end{equation}
The other singlet is $|s_{4,b}\rangle \equiv 
|s_2\rangle^{\otimes 2}$, 
i.e., two
copies of the second order singlet $|s_2\rangle$. It can be written explicitly as
\begin{eqnarray}
|s_{4,b}\rangle 
&=&
\frac{1}{2}\epsilon_{ij}\epsilon_{kl}
\left|
\begin{array}{c}
i \\ j \\ k\\ l
\end{array}
\right\rangle \\
&=&
\frac{1}{2}
\left(
\left|
\begin{array}{c}
0 \\ 1 \\ 0\\ 1
\end{array}
\right\rangle
+
\left|
\begin{array}{c}
1 \\ 0 \\ 1\\ 0
\end{array}
\right\rangle
-
\left|
\begin{array}{c}
0 \\ 1 \\ 1\\ 0
\end{array}
\right\rangle
-
\left|
\begin{array}{c}
1 \\ 0 \\ 0\\ 1
\end{array}
\right\rangle
\right).
\end{eqnarray}

We can make entanglement monotones by using these singlets. For example,
$|s_{4,a}\rangle \otimes |s_{4,a}\rangle \otimes |s_{4,a}\rangle$
is represented as
\begin{eqnarray}
|s_{4,a}\rangle \otimes |s_{4,a}\rangle \otimes |s_{4,a}\rangle
&=&
d_{i_{11}i_{21}i_{31}i_{41}} 
d_{i_{12}i_{22}i_{32}i_{42}} 
d_{i_{13}i_{23}i_{33}i_{43}}
\left|
\begin{array}{ccc}
i_{11} & i_{12} & i_{13} \\
i_{21} & i_{22} & i_{23} \\
i_{31} & i_{32} & i_{33} \\
i_{41} & i_{42} & i_{43}
\end{array}
\right\rangle, 
\end{eqnarray}
and
\begin{equation}
|\psi^{\otimes 4}\rangle 
=
c_{i_{11}i_{12}i_{13}}
c_{i_{21}i_{22}i_{23}}
c_{i_{31}i_{32}i_{33}}
c_{i_{41}i_{42}i_{43}}
\left|
\begin{array}{ccc}
i_{11} & i_{12} & i_{13} \\
i_{21} & i_{22} & i_{23} \\
i_{31} & i_{32} & i_{33} \\
i_{41} & i_{42} & i_{43}
\end{array}
\right\rangle, 
\end{equation}
where $|\psi\rangle = c_{ijk}|ijk\rangle$ is a 3-qubit state. 
Then we obtain
\begin{eqnarray}
&&\left| \left(\langle s_{4,a}| \otimes \langle s_{4,a}| \otimes \langle s_{4,a}|\right)
 | \psi^{\otimes 4}\rangle 
\right| 
\\
&=& 
\left| d_{i_{11}i_{21}i_{31}i_{41}} d_{i_{12}i_{22}i_{32}i_{42}} d_{i_{13}i_{23}i_{33}i_{43}} 
c_{i_{11}i_{12}i_{13}} c_{i_{21}i_{22}i_{23}} c_{i_{31}i_{32}i_{33}} c_{i_{41}i_{42}i_{43}} \right|.
\end{eqnarray}
By explicit calculation it is shown to be $\frac{\tau_3}{8\sqrt{3}}$, where $\tau_3$ is the 3-tangle
\cite{3tangle}.

There are other fourth order singlets: 
$|s_{4,b}\rangle\otimes |s_{4,a}\rangle \otimes |s_{4,a}\rangle$,
$|s_{4,b}\rangle\otimes |s_{4,b}\rangle \otimes |s_{4,a}\rangle$,
$|s_{4,b}\rangle\otimes |s_{4,b}\rangle \otimes |s_{4,b}\rangle$
and their permutations.  
However, we have to use singlets with even number of $|s_{4,b}\rangle$,
since 
$|s_{4,b}\rangle$ is antisymmetric with respect to the exchange of a pair
of copies of $|\psi\rangle$. Therefore we use 
$|s_{4,b}\rangle\otimes |s_{4,b}\rangle \otimes |s_{4,a}\rangle$,
and obtain
\begin{eqnarray} 
&&
(\langle s_{4,b}|\otimes \langle s_{4,b}| \otimes \langle s_{4,a}|)
|\psi^{\otimes 4}\rangle \\
&=& 
\frac{1}{4}
\left|
\epsilon_{i_{11}i_{21}}\epsilon_{i_{31}i_{41}}
\epsilon_{i_{12}i_{22}}\epsilon_{i_{32}i_{42}}
d_{i_{13}i_{23}i_{33}i_{43}}
c_{i_{11}i_{12}i_{13}} c_{i_{21}i_{22}i_{23}} c_{i_{31}i_{32}i_{33}} c_{i_{41}i_{42}i_{43}}
\right|
\\
&=&
\frac{\tau_3}{8\sqrt{3}}.
\end{eqnarray}
Considering the permutation symmetry of $\tau_3$,
we see that $\tau_3$ is the only fourth-order 
invariant polynomial in the 3-qubit case.

It is possible to choose another basis for the fourth order singlets.
For example, let us make two second order singlets by combining
the first and third copies, and the second and the fourth
ones. By putting together the two second order singlets
we obtain another fourth order singlet
\begin{eqnarray}
|s_{4,c}\rangle
&\equiv&
\frac{1}{2}\epsilon_{ik}\epsilon_{jl}
\left|
\begin{array}{c}
i \\ j \\ k\\ l
\end{array}
\right\rangle \\
&=&
\frac{1}{2}
\left(
\left|
\begin{array}{c}
0 \\ 0 \\ 1\\ 1
\end{array}
\right\rangle
+
\left|
\begin{array}{c}
1 \\ 1 \\ 0\\ 0
\end{array}
\right\rangle
-
\left|
\begin{array}{c}
1 \\ 0 \\ 0\\ 1
\end{array}
\right\rangle
-
\left|
\begin{array}{c}
0 \\ 1 \\ 1\\ 0
\end{array}
\right\rangle
\right).
\end{eqnarray}
It is linearly dependent on $|s_{4,a}\rangle$ and $|s_{4,b}\rangle$:
$|s_{4,c}\rangle = \frac{\sqrt{3}}{2}|s_{4,a}\rangle + \frac{1}{2}|s_{4,b}\rangle$.
Therefore it does not give any new measure, but it gives new expressions for $\tau_3$. 
For example, 
\begin{eqnarray}
&&
(\langle s_{4,b}|\otimes \langle s_{4,b}| \otimes \langle s_{4,c}|)
|\psi^{\otimes 4}\rangle \\
&=& 
\frac{1}{8}
\left|
\epsilon_{i_{11}i_{21}}\epsilon_{i_{31}i_{41}}
\epsilon_{i_{12}i_{22}}\epsilon_{i_{32}i_{42}}
\epsilon_{i_{13}i_{33}}\epsilon_{i_{23}i_{43}}
c_{i_{11}i_{12}i_{13}} c_{i_{21}i_{22}i_{23}} c_{i_{31}i_{32}i_{33}} c_{i_{41}i_{42}i_{43}}
\right|
\\
&=&
\frac{\tau_3}{16}
\label{3tangle_original}
\end{eqnarray}
is the original form of the 3-tangle given in \cite{3tangle}.

\section{Other cases}
We can construct entanglement monotones for multiqubit systems in the same way.
The simplest singlet is $|s_2\rangle\otimes\dots\otimes |s_2\rangle$.
The corresponding entanglement measure is 
\begin{eqnarray}
\frac{1}{2^{m/2}}
\left|
\epsilon_{i_{11}i_{21}}\, 
\epsilon_{i_{12}i_{22}}\dots  
\epsilon_{i_{1m}i_{2m}}\,
c_{i_{11}i_{12}\dots i_{1m}}\, 
c_{i_{21}i_{22}\dots i_{2m}}
\right|,
\end{eqnarray}
which is a generalization of the concurrence. 
Although this second order measure vanishes identically for odd $m$,
we can construct fourth order measures using $|s_{4,a}\rangle$ and $|s_{4,b}\rangle$. 
For example, the measure corresponding to $|s_{4,a}\rangle \otimes \dots \otimes |s_{4,a}\rangle$ is
\begin{eqnarray}
&&
\left|d_{i_{11}i_{21}i_{31}i_{41}}\dots d_{i_{1m}i_{2m}i_{3m}i_{4m}}
c_{i_{11}\dots i_{1m}} c_{i_{21}\dots i_{2m}} 
c_{i_{31}\dots i_{3m}} c_{i_{41}\dots i_{4m}} \right|,
\end{eqnarray}
which is a natural generalization of the 3-tangle.
It vanishes for the $m$-qubit W state $|W_m\rangle$,
since $|W_m\rangle^{\otimes 4}$ is an eigenstate of
the $z$ component of the total spin with eigenvalue
$2m$, and hence has no singlet component.
For the $m$-qubit GHZ state, it takes the value
$\left\{2^{m+1} + 4 (-1)^m\right\}/(2\sqrt{3})^m$.
Unlike the 3-qubit case, there are many other fourth order invariants for $m\ge 4$. 

It is also possible to construct entanglement measures for multilevel systems. 
As a simple example, let us consider a pair of qutrits.
The simplest ${\rm SU}(3)$ singlet which can be constructed from qutrits is the third order 
completely antisymmetric state 
\begin{eqnarray}
|s_3\rangle &\equiv& \frac{1}{\sqrt{6}} \epsilon_{ijk}
\left|
\begin{array}{c}
i \\ j \\ k
\end{array}
\right\rangle .
\end{eqnarray}
Note that the indices take on one of the three values 
$0$, $1$, and $2$ in this case.
Then the measure corresponding to $|s_3\rangle\otimes |s_3\rangle$ is
\begin{eqnarray}
\frac{1}{6}
\left|\epsilon_{i_{11}i_{21}i_{31}} \epsilon_{i_{12}i_{22}i_{32}} 
c_{i_{11}i_{12}}c_{i_{21}i_{22}}c_{i_{31}i_{32}}\right|.
\end{eqnarray}
This is equal to $|\det C|$, where $C\equiv \{c_{ij}\}$ is a $3\times 3$ matrix.
It is obvious how to generalize this measure to a $N\times N$ system. 

In general, a singlet representation of ${\rm SU}(N)$ can be constructed
from $kN$ defining representations, where $k\ge 1$ is an integer. 
For $k=1$, the completely antisymmetric representation is the only singlet.
There are many singlets for $k\ge 2$. For example, there are
three sixth order singlets for ${\rm SU}(3)$.

\section{Nonsinglet representations}
A representation whose highest weight is neither "maximum" nor "minimum"
(i.e., $\bnu_{\alpha} \ne q\be, \bzero$) also seems useful to classify entanglement,
since the projection thereto vanishes for nonentangled states.
We do not know if its norm (or some function thereof) is an entanglement monotone or not.
We cannot apply our previous proof of monotonicity in this case
because the SLOCC group does not preserve the norm and hence
$|P_{\bnu_\alpha}|\psi^{\otimes q}\rangle |$ is not an invariant
of that group unless $\bnu_\alpha = 0$.
We can, however, show that if $P_{\bnu_\alpha}|\psi^{\otimes q}\rangle \ne 0$
and $P_{\bnu_\alpha}|\phi^{\otimes q}\rangle = 0$,
$|\psi\rangle$ and $|\phi\rangle$ does not belong to the same SLOCC class.   
The point is that an irrep $R_{\bnu_\alpha}$ of the LU group is also
an irrep of the SLOCC group, since the Lie algebra of the SLOCC group is 
the complexification of the Lie algebra of the LU group. Therefore 
$g P_{\bnu_\alpha} |\psi^{\otimes q}\rangle = P_{\bnu_\alpha}\left(g|\psi\rangle\right)^{\otimes q} 
\in R_{\bnu_\alpha}$ 
for any $g$ in the SLOCC group, and if 
$P_{\bnu_\alpha}|\psi^{\otimes q}\rangle \ne 0$, 
$P_{\bnu_\alpha}\left(g|\psi\rangle\right)^{\otimes q} \ne 0$
because $g$ is invertible. 
Hence there is no $g$ in the SLOCC group such that $g|\psi\rangle = |\phi\rangle$.

Let us consider two copies of a state in a 
$N_1\times\dots\times N_m$ multipartite system
as an example.
Two copies of a defining representation of ${\rm SU}(N_i)$ is
decomposed into symmetric and antisymmetric irreducible components.
We denote the projection operators for the symmetric and antisymmetric 
components 
by $P_{+,i}$ and $P_{-,i}$, respectively. 
The norm of an irreducible component
$| P_{s_1,1}\otimes \dots \otimes P_{s_m,m} |\psi^{\otimes 2}\rangle |\;
(s_j = \pm) $ is 
the generalized concurrence defined in \cite{concurrence2}.
For example, for the 3-qubit case \cite{sugita1}
\begin{eqnarray}
|P_{-,1}\otimes P_{-,2}\otimes P_{+,3}|\psi^{\otimes 2}\rangle|^2
&=& 
\frac{1}{4}C_{AB}^2 + \frac{1}{8}\tau_{3}.
\label{3qubit_conc}
\end{eqnarray} 
Here, $C_{AB}$ is the concurrence \cite{wootters} for the
first and the second qubits.

\section{Experimental observation}
Since an irrep of the LU group can be written in the product form $R_{\nu_1}\otimes \dots \otimes R_{\nu_m}$,
the projection operator thereto can also be written as the tensor product of 
local projection operators $P_{\bnu} = P_{\nu_1}\otimes \dots \otimes P_{\nu_m}$
\footnote{If there are more than one independent singlets of the LU group 
of a given order, 
it is possible to
make another singlet from a linear combination of the singlets. 
For example, in the 4-qubit case, we can make a 4th order singlet
of the form 
$|s_{4,a}\rangle \otimes |s_{4,b}\rangle \otimes |s_{4,b}\rangle \otimes |s_{4,b}\rangle +
|s_{4,a}\rangle \otimes |s_{4,a}\rangle \otimes |s_{4,b}\rangle \otimes |s_{4,b}\rangle$.
In such a case, the corresponding projection operator
cannot be written as a product of local projection operators.
Then we have to combine several different local measurements to determine the value of the
corresponding measure.}
.
Therefore the value of our measure can be determined by local projective measurements
of some copies of the state.
Note that a recently proposed scheme for experimental determination of the SLOCC class
of a 3-qubit state \cite{yu_song}
is easily obtained from our method using  
Eqs. (\ref{3tangle_original}) and (\ref{3qubit_conc}).

\section{Conclusion and remarks}
In this paper, we have considered irreducible decomposition of $q$ copies
of a quantum state $|\psi^{\otimes q}\rangle$ with respect to the
LU group, and shown that the norm of the projection
to an irreducible component can be a good measure of entanglement.
The irreps can be classified into three types:
(A) irrep with the "maximum" highest weight, (B) irrep with
the "minimum" highest weight (singlet), (C) others.
If the state $|\psi\rangle$ is unentangled, $|\psi^{\otimes q}\rangle$
is in the type A irrep. Therefore projections
to type B and type C components vanish. We have shown that 
an entanglement monotone is obtained from  
the projection to a type B component. We have also shown that
the projection to a type C component is useful for 
SLOCC classification of entanglement.

A remaining problem of our method is that independent irreducible 
components do not necessarily give independent measures.
For example, there are many (actually $2^3=8$) independent fourth order singlets in the 
3-qubit case, but the 3-tangle is the only measure obtained from them.  
It happens because we do not take into account the permutation symmetry
of the copies of a quantum state; since $|\psi^{\otimes q}\rangle$ is completely
symmetric with respect to permutations of the copies, 
only completely symmetric irreducible components survive after projection.
It means that we should consider irreducible decomposition not only
under the local unitary group, but also under
the permutation group of the copies.
Progress in this direction will be reported in the forthcoming paper
\cite{sugita3}. 
The permutation group of the parties also would be useful.
We hope extensive use of the group theory would provide insight
into the nature of multipartite entanglement.

Another important problem is to generalize our method from pure to mixed states.
We can formally define an entanglement measure for mixed states 
$E(\rho)$ from that for pure states $E(|\psi\rangle )$ via the convex roof
construction \cite{bennet, uhlmann}
\begin{eqnarray}
E(\rho) = {\rm min} \sum_{i} p_i E(|\psi_i\rangle ), \;\;\;\; 
\rho = \sum_i p_i |\psi_i\rangle \langle \psi_i|.
\end{eqnarray} 
Although it is in general difficult to find the set $\{p_i\}$ which gives 
the minimum value, the group theoretical nature of our method and the projection
operator form of the measures could help in solving this problem.

\begin{acknowledgements}
The author would like to thank A. Miyake for
helpful comments.
\end{acknowledgements}


\begin{thebibliography}{99}


\bibitem{3qubit}
W. D\"ur, G. Vidal, and J.I. Cirac,
Phys. Rev. A {\bf 62}, 062314 (2000).

\bibitem{miyake}
A. Miyake,
Phys. Rev. A {\bf 67}, 012108 (2003).

\bibitem{sugita1}
A. Sugita, 
J. Phys. A {\bf 36}, 9081 (2003).


\bibitem{concurrence2}
F. Mintert,  M. Ku\'s, and A. Buchleitner,
Phys. Rev. Lett. {\bf 95}, 260502 (2005).

\bibitem{osterloh}
A. Osterloh and J. Siewert, Phys. Rev. A {\bf 72}, 012337 (2005); Int. J. Quant. Inf. 4, 531 (2006).

\bibitem{georgi}
H. Georgi, {\it Lie Algebras in Particle Physics} (Benjamin/Cumming, Reading, MA, 1982).


\bibitem{sugita2}
A. Sugita, J. Phys. A: Math. Gen. {\bf 35}, L621 (2002).

\bibitem{mintert_zyczkowski}
F. Mintert and K. \.Zyczkowski, Phys. Rev. A {\bf 69}, 022317 (2004).

\bibitem{verstraete}
F. Verstraete, J. Dehaene, and B. De Moor, Phys. Rev. A {\bf 68},
012103 (2003).

\bibitem{gert}
S. Schenk and G.-L. Ingold,
Phys. Rev. A {\bf 75}, 022328 (2007).

\bibitem{3tangle}
V. Coffman, J. Kundu, and W. K. Wootters,
Phys. Rev. A {\bf 61}, 052306 (2000).


 



\bibitem{wootters}
W. K. Wootters, Phys. Rev. Lett. {\bf 80}, 2245 (1998).



\bibitem{yu_song}
C. S. Yu and H. S. Song, Phys. Rev. A {\bf 76}, 022324 (2007).

\bibitem{sugita3}
A. Sugita, in progress.

\bibitem{bennet}
C. H. Bennett, D. P. DiVincenzo, J. A. Smolin, and W. K.
Wootters, Phys. Rev. A {\bf 54}, 3824 (1996).

\bibitem{uhlmann}
A. Uhlmann, Phys. Rev. A {\bf 62}, 032307 (2000).

\end{thebibliography}
\end{document}